\documentclass{aa}
\usepackage{graphicx}
\usepackage{natbib}
\begin{document}
   \title{First detection of the growing humps at the rapidly rising stage
   of dwarf novae AL Com and WZ Sge}
   \titlerunning{First detection of the growing humps}

\author{Ryoko Ishioka\inst{1}
     \and      Makoto Uemura\inst{1}
     \and      Katsura Matsumoto\inst{1,2}
     \and      Hiroyuki Ohashi\inst{1}
     \and      Taichi Kato\inst{1}
     \and      Gianluca Masi\inst{3} 
     \and      Rudolf Novak\inst{3}
     \and      Jochen Pietz\inst{3}
     \and      Brian Martin\inst{3}
     \and      Donn Starkey\inst{3}
     \and      Seiichiro Kiyota\inst{3}
     \and      Arto Oksanen\inst{3}
     \and      Marko Moilanen\inst{3}
     \and      Lew Cook\inst{3}
     \and      Lukas Kral\inst{3}
     \and      Tomas Hynek\inst{3}
     \and      Marek Kolasa\inst{3}
     \and      Tonny Vanmunster\inst{3}
     \and      Michael Richmond\inst{3}
     \and      Jim Kern\inst{3}
     \and      Stacey Davis\inst{3}
     \and      Dustin Crabtree\inst{3} 
     \and      Kevin Beaulieu\inst{3}
     \and      Tracy Davis\inst{3} 
     \and      Matt Aggleton\inst{3}
     \and      Kosmas Gazeas\inst{3}
     \and      Panos Niarchos\inst{3}
     \and      A. Yushchenko\inst{3}
     \and      Franco Mallia\inst{3} 
     \and      Marco Fiaschi\inst{3}
     \and      Gerry A. Good\inst{3}
     \and      David Boyd\inst{3} 
     \and      Yasuo Sano\inst{3}
     \and      Koichi Morikawa\inst{3}  
     \and      Masayuki Moriyama\inst{3}
     \and      Ronald Mennickent\inst{3}
     \and      Jose Arenas\inst{3}
     \and      Tomohito Ohshima\inst{3}
     \and      Tutomu Watanabe\inst{3}}


\institute{Department of Astronomy, Faculty of Science,
                 Kyoto University, Kyoto 606-8502, Japan
\and Graduate School of Natural Science and Technology, 
                 Okayama University, Okayama 700-8530, Japan
\and VSNET Collaboration Team}

   \date{Received October 29, 2001; accepted November 21, 2001}

   \abstract{
    We report on time-series photometric observations in the earliest stages 
 of  superoutbursts of the extreme dwarf novae, AL Com and WZ Sge, which
 started on 2001 May after the 6 years quiescence and on 2001 July after
 the 23 years quiescence, respectively. We detected growth of ``early 
 superhumps'' during the each rising stage. Our observations reject the mass
 transfer instability for the trigger of the superoutburst of WZ Sge
 stars, and show the existence of some relations between the ``early
 superhumps'' and the spiral structure, which give a hint of the origin
 of the ``early superhumps.''
   }

\maketitle

\section{Introduction}

Dwarf novae are cataclysmic binaries containing a Roche lobe-filling
dwarf star and a white dwarf with an accretion disk (for a review, see
\citealt{war95book}). SU UMa-type dwarf novae (SU UMa stars) are a subclass of
dwarf novae, which are characterized by two types of outbursts - short
``normal'' outbursts and long ``superoutbursts''. During superoutbursts,
periodic modulations called ``superhumps'' are observed, 
whose period is slightly longer than the orbital period. The best accepted
mechanism for SU UMa stars is a combination of thermal and tidal
instabilities in the accretion disk (DI model; \citealt{osa89suuma}), and the
superhump phenomenon is well explained by the tidally distorted accretion
disk (\citealt{ich93SHmasstransferburst}).  

WZ Sge-type dwarf novae (originally proposed by \citealt{bai79wzsge}, and
\citealt{kat01hvvir}) form an 
extreme subgroup of SU UMa stars. The well-established members are only
5 stars, HV Vir, AL Com, EG Cnc, RZ Leo, and WZ Sge itself. They also
show superoutbursts and superhumps, but have many peculiar respects
different from {\it usual} SU UMa stars. The outburst frequency of WZ Sge
stars is very low and the interval of two successive outbursts
(supercycle) is years to decades ({\it usually} months), the outburst
amplitude is more than 6 mag ({\it usually} 2-5 mag), the outburst duration is
months ({\it usually} two weeks), and no or few normal outburst occurs in one
supercycle ({\it usually} several ones).  

However, these characteristics are relative ones, and there has been a
suggestion that WZ Sge stars and SU UMa stars comprise a continuous
entity, rather than a separate class (\citealt{osa95wzsge}).
The only diagnostic characteristics is the existence of ``early
superhumps.''\footnote{The same feature is also referred to as {\it orbital
superhumps} (\citealt{kat96alcom}) or {\it outburst orbital hump}
(\citealt{pat98egcnc}).  We use the term in this paper ``early superhumps''
because the period of this feature was shown to be slightly, but significantly
different from the orbital period (section 2), qualifying the 
{\it superhump} nature of the signal, and because we consider that the
origin of the feature is tidal resonance, rather than an enhanced hot spot
(section 3).}
Early superhumps are periodic modulations with a period  
of identical to the orbital period and often with a complicated profile,
which is observed in the earliest stage of the superoutburst of the WZ Sge
stars. The early superhumps are observed only in the WZ Sge stars, 
but never in the other SU UMa stars. Thus the understanding of early
superhumps is very important in the understanding of WZ Sge stars as a
distinct group, but the origin is still an open question. 

Many authors have tried to explain the origin of the peculiarity
of the WZ Sge stars. \citet{osa95wzsge} showed that if the very low mass
transfer rate and the extremely low viscosity parameter, $\alpha$, of
the quiescent disk is assumed, the long cycle length and the large scale
outburst can be explained  in the disk instability model. Another
explanation is a mass transfer burst model, which explains the outburst
by the sudden increase of the mass transfer rate from the secondary
(\citealt{pat81wzsge}). 

The judgment on the two models is given by the observations of the
outburst onset. However, the difficulty of the observational study of WZ
Sge stars is in their long supercycle length (e.g. 32.5 years of WZ Sge),
as well as, the shortness of the rising stage of the outburst of dwarf
novae. During the last outburst of WZ Sge in 1978, unfavorable seasonal
conditions made detailed and continuous photometric observations difficult. Two
superoutbursts of AL Com in 1995 and EG Cnc in 1997 were rather well
observed (\citealt{kat96alcom};\citealt{nog97alcom};\citealt{mat98egcnc}, and
references therein). However, the rising stages of the outbursts were
not covered by the observations in both outbursts, too.
 Many theoretical studies have been done based on the relatively fragmentary
data of WZ Sge during previous outbursts and the characteristics of the
other members of the WZ Sge stars, but more detailed data of WZ Sge stars
during outbursts are necessary to test the theories. So the further
outbursts of WZ Sge stars, and the precise observations, 
especially the observations of the earliest stage, have been strongly
desired for a long time.  

Here, we report on the first detection of the growth of periodic
modulations in the earliest stage of the long-awaited
two superoutbursts of AL Com and WZ Sge, which were discovered by  S. Kerr
on 2001 May 18.51 at $m_{\rm vis}=$13.4 and by T. Ohshima on 2001 July
23.565 UT at $m_{\rm vis}=$ 9.7, respectively. For both outbursts,
receiving the reports of the outbursts, we immediately started extensive
international campaign of time-series photometric observations through
the Variable Star NETwork (VSNET, {\tt
http://www.kusastro.kyoto-u.ac.jp/vsnet/}). Both campaigns succeeded in
detecting the rising stages of the outbursts and revealed that 
periodic modulations (early superhumps) grew after the outburst onset. In this
{\it letter}, we focus our interest on the earliest stages of the
outbursts. The 
details of the observations of WZ Sge and AL Com will be issued in
forthcoming papers. 

\section{Observations and results}

The CCD observations reported here were performed by the international
VSNET collaboration team using 5.5--80~cm telescopes with  time
resolutions of 0.5--100~sec. 

The magnitudes of AL Com and WZ Sge were measured using several
comparison stars (for WZ Sge, see \citealt{hen01wzsge}), and
adjusted to a common scale close to the $R_{\rm c}$ system by correcting
for the systematic difference between observers. 
Heliocentric corrections to the observed times were applied before the 
following analyses. 

Figure 1 shows the light curves by the earliest observations of the two
outbursts. Our observations detected the rapid brightenings with a rate of
more than 3 mag d$^{-1}$ for AL Com and more than 5 mag d$^{-1}$ for WZ 
Sge. In both outbursts, no clear modulation was observed just after the
outburst onset, but the full amplitude modulations grew up before the
outburst maxima. The light curve of WZ Sge clearly shows the growth   
of periodic modulations superimposed on the rising trend. The growing time
scale of the periodic modulations was only one day.

Figure 2 shows the the whole light curves of the outbursts. Since the
presence of two types of superhumps in AL Com during its 1995 
superoutburst was already described in the literature (\citealt{kat96alcom}),
we mainly focus on WZ Sge in this {\it letter},
especially on the clarification of two types of humps.

As shown in the upper panel of figure 2, the main outburst continued
for 23 days. The declining rate was more than 0.2 mag d$^{-1}$ at first,
however, finally settled into 0.13 mag d$^{-1}$ which is a typical
value for the SU UMa stars. 

We measured the peak timings of the humps during the main outburst and
performed a period analysis. The cycle counts start by $E=0$ at HJD
2452115.0330. There is no uncertainty concerning our cycle count because
of quite dense observations. The typical error of the timing is 0.001~d.
In the case that the hump is 
doubly-peaked, the timings of both peaks were measured for one cycle
count. Figure 3 shows the $O-C$ diagram 
calculated for the orbital period of 0.05668784707~d (\citealt{ski98wzsge}).  

During the period of July 23 -- August 3 ($0 \leq E \leq 206$), the major peak
timing of recurring humps with a stable period is best fitted by the
following equation:  
  $Peak(HJD) = 0.056656(2)E + 2452115.0000(2)$.
The period of 0.056656(2)~d is 0.05~\% shorter than the orbital period of
0.05668784707~d \citep{ski98wzsge}, but the two periods are very close to each
other. On August 4, a different type of humps with a longer period
appeared. During the period of August 4 -- 16 ($206 \leq E \leq 407$), the peak
timing is fitted by another equation: 
  $Peak(HJD) = 0.05726(1)E + 2452114.890(4)$.
The period of 0.05726(1)~d is 1.01~\% longer than the orbital period.
These two periods were also confirmed by a period analysis using PDM
method.

Figure 4 shows the averaged light curves of two types of
humps. The former type of humps with a period of 0.056656(2)~d,
almost identical to the orbital period, are ``early superhumps.'' The
latter type of humps with a period of 0.05726(1)~d, 1.0~\% 
longer than the orbital period, are ``superhumps.'' 

  Although there had been a suggestion of two types of periodic
modulation during the 1978 outburst (\citealt{boh79wzsge}), the
evidence was not yet compelling because of the remaining
ambiguity in identifying the hump maxima and cycle counts, as already
discussed by \citet{pat81wzsge}. Our continuous coverage of the
entire outburst first unambiguously established the two periods of 
0.056656(2)~d and 0.05726(1)~d for ``early superhumps'' and ``genuine''
superhumps, respectively\footnote{ The superhump period of 0.05714~d and
the fractional superhump excess of 0.8~\% reported by \citet{pat81wzsge} were
found to be incorrect, which suffered from ambiguous 
identifications of superhumps.}.

The main outburst terminated on August 16,
but the rapid fading stopped on August 18 at 2 mag brighter than its
quiescence level. After the dip for 5 days, twelve repetitive
rebrightenings were observed between August 21 and September 13. This
occurrence of rebrightenings far surpassed the past record of 
six rebrightenings in EG Cnc.

\section{Discussion}

The detection of the growth of early superhumps ``after'' the outbursts
onset of AL Com and WZ Sge puts a period to the long-continuing
argument between the competing two models; 
the mass-transfer burst model (\citealt{pat81wzsge}) and the disk-instability
model (\citealt{osa95wzsge}), originally continuing since 1970s. 
Although several different theoretical interpretations have been proposed
as the cause of WZ Sge outbursts, no observational test has been available
before the present outburst.  We can now present the {\it first} firm
observational evidence against the {\it original} mass-transfer burst model
for the cause of WZ Sge outburst, proposed by \cite{pat81wzsge}. If the 
superoutburst is triggered by an enhanced mass-transfer, much larger
humps due to the brightened hot spot are expected at the same time
with the outburst onset. This is clearly contrary to our observation.
This is also supported by Doppler tomography of the earliest stage of the
superoutburst (Baba et al. 2001).  The strongest emission on velocity
maps reside on the opposite side of the expected hot spot, excluding an
increased mass-transfer as the origin of the strong emissions.

Some mechanisms for the early superhumps have been proposed by
several authors.  Our double-peaked profile (figure 4)
excludes the hot spot model, since the profile would be single-peaked
if the hot spot was the origin (\citealt{kat96alcom}). \cite{nog97alcom}
proposed a
jet model and a thickened-edge disk model. Recently, a possibility, that
the disk reaches to the 3:1 resonance radius before the outburst, is
pointed out \citep{mey98wzsge}, and we thought that the slightly distorted
disk might be the origin. However, no consensus has been yet reached as to
the origin of early superhumps.

Osaki \& Meyer (2001; in prep.) suggest that the double peaked light
curve at the early stage of the outburst of WZ Sge stars is a
manifestation of the 2:1 resonance at the accretion disk rim that was
first investigate by \cite{lin79}.
In binary systems with an extremely small mass ratio, the
tidal force is very weak, and the accretion disk can extend so large
that a strong 2:1 resonance begins to work (\citealt{lin79}). A spiral
dissipation pattern is produced by this resonance, which may draw a
double-peaked light curve. The spiral pattern formed in the extended
disk at the outburst onset may be a natural explanation of our
observations.

We compared our photometric observations with the spectroscopic
observations of WZ Sge. On July 23, just after the outburst onset, we
found no emission component at the place of HeII 4686, but on July 24,
the line was observed as
a strong emission line. The Doppler maps of \ion{He}{ii} $\lambda4686$
constructed using 
time-resolved spectra on July 24 and July 28 show that the
accretion-disk emission is dominated by two spiral arms
(\citealt{ish01wzsgeiauc7669};\citealt{ste01wzsgeiauc7675},
\citealt{bab01wzsgeiauc7672};\citealt{bab01wzsgeiauc7678}). This 
result indicates that some 
relation exists between the growth of \ion{He}{ii} emission line and the spiral
structure, and the observed growth of early superhumps. The
double-peaked profile of early superhumps may be explained by the
two arms of the spiral structure.\footnote{Spiral shocks are seen in
other cataclysmic variables (e.g. IP Peg) that do not show similar
humps. Also, since IP 
Peg has a long orbital period, neither the 2:1 nor the 3:1 resonances
will occur in that star. However, the mechanisms driving spiral shocks
of WZ Sge and above-the-gap dwarf novae can be different. (Baba et al. 2001)}

We propose that the physical reason, distinguishing the WZ Sge stars
from other dwarf novae, is a very small mass ratio which
allows the existence of the 2:1 resonance. The condition of the 2:1
resonance being inside the Roche-lobe is the mass ratio of
$q = M_{2}/M_{1} \leq 0.1$. The mass ratio is indicated from the
superhump excess, 
$\epsilon$, which is expressed as $\epsilon=(P_{\rm sh}-P_{\rm
orb})/P_{\rm orb}$. WZ Sge stars clearly have small
mass ratios compared with usual SU UMa stars. There are only several
stars containing WZ Sge stars in the region where the 2:1 resonance can
be effective. We succeeded in understanding the WZ Sge
stars in the unified way by a single mechanism.  

Our observations with high time-resolution and long, continuous coverage
not only revealed the behavior at the earliest stage of the outburst,
rejecting the mass transfer instability model, but also lead the
condition of $q\leq$ 0.1, 
distinguishing WZ Sge stars from SU UMa stars. WZ Sge stars 
resemble X-ray transients containing black holes in some aspects.  The
investigation of WZ Sge stars is expected to contribute the
understanding of the black hole transients with a similar mass ratio. 

\begin{acknowledgements}
The international observing campaign through VSNET, totally
unprecedented in its extent and temporal coverage made possible the
immediate start of the photometric and spectroscopic observations after
the outburst detection, and this dense and outstandingly long light
curve presented here. We are grateful to numerous collaborators of VSNET
for the large contribution to this observation.  We thank T. Ohshima for
the discovery of the outburst at the earliest stage. We acknowledge
Y. Osaki and D. Nogami for the fruitful discussion and useful advice. 
Part of this work was supported by a Research Fellowship of the Japan Society 
for the Promotion of Science for Young Scientists (M.U.).
\end{acknowledgements}

\begin{figure}
   \resizebox{\hsize}{!}{\includegraphics{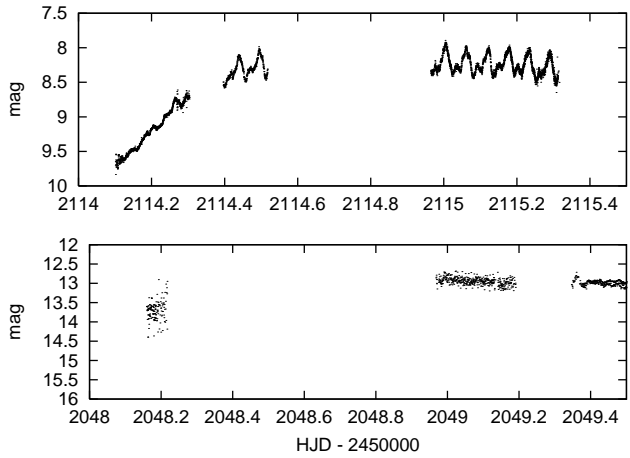}}
  \caption{Light curves of the earliest stage of the 2001 superoutbursts
 in AL Com (lower panel) and WZ Sge (upper panel). In the case of AL Com, no clear hump was observed 
 in the rising stage, although a hint of hump appeared. In the case of
 WZ Sge, the growth of humps was clearly detected during the rising stage.}
\label{fig:figure1}
\end{figure}

\begin{figure}
   \resizebox{\hsize}{!}{\includegraphics{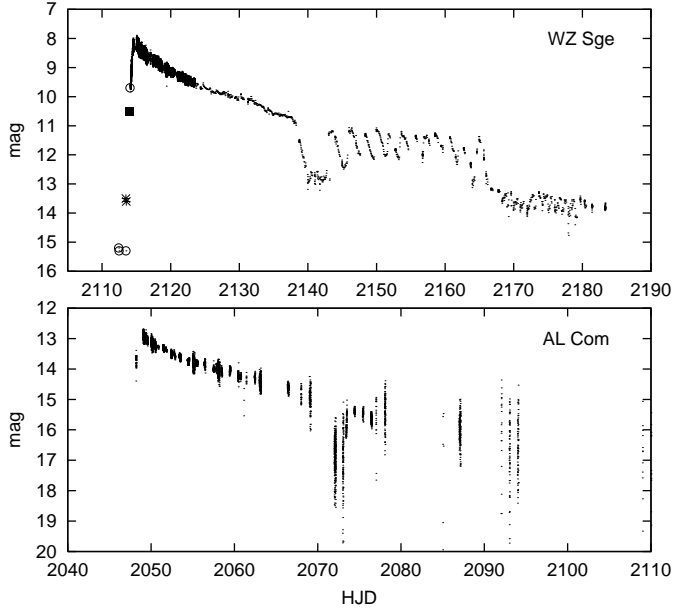}}
  \caption{General light curves of the 2001 superoutbursts in AL Com
 (lower panel) and WZ
 Sge (upper panel). Small dots represent our CCD observations, filled square
 represents photographic observation, open circles represent visual
 observations, and crosses represent negative observations.}
\label{fig:figure1}
\end{figure}

\begin{figure}
   \resizebox{\hsize}{!}{\includegraphics{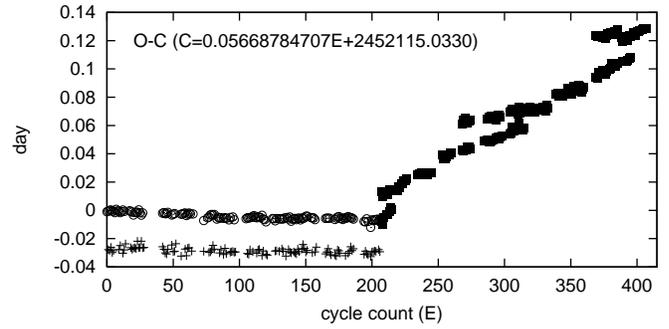}}
  \caption{The $O-C$ diagram of the hump-maxima timings. open circles
 represent major peaks of early superhumps, cross represent minor
 peaks of early superhumps, and filled squares represent peaks of superhumps.}
\label{fig:figure2}
\end{figure}

\begin{figure}
  \resizebox{\hsize}{!}{\includegraphics{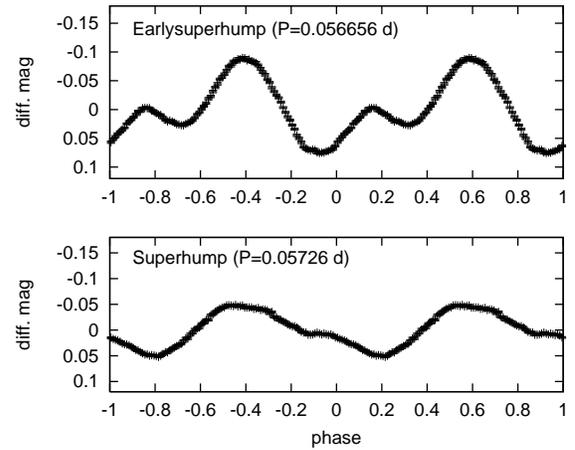}}
  \caption{Averaged light curves of early superhumps and superhumps of
 WZ Sge. Upper panel shows an averaged light curve of early superhumps
 folded by a period of 0.056656~d. Lower panel shows an averaged light
 curve of superhumps folded by a period of 0.05726~d.}
\label{fig:figure3}
\end{figure}

\end{document}